\pdfoutput=1
\documentclass[
twocolumn,
english,
aps,
prb,
groupedaddress,
showpacs,
twocolum,
superscriptaddress,
nobibnotes
]{revtex4-1}
\usepackage[T1]{fontenc}
\usepackage[latin9]{inputenc}
\usepackage{geometry}
\usepackage{verbatim}
\usepackage{amsmath}
\usepackage{amssymb}
\usepackage{esint}
\usepackage[nodisplayskipstretch]{setspace}
\usepackage{hyperref}
\usepackage[hyphenbreaks]{breakurl}

\makeatletter
\usepackage{caption}
\usepackage{subcaption}
\usepackage{wrapfig}
\usepackage{braket}
\usepackage{array}
\usepackage{relsize}
\usepackage{indentfirst}
\usepackage{simplewick}
\usepackage{mathrsfs}

\usepackage{newtxmath}
\usepackage{newtxtext}
\usepackage{mleftright}
\mleftright

\renewcommand{\[}{\begin{equation}}
    \renewcommand{\]}{\end{equation}}

\usepackage{lipsum}
\def\Eq#1{Eq. {\eqref{#1}}}

\makeatother

\usepackage{babel}
\begin{document}
% \title{The inner workings of fractional quantum Hall parent Hamiltonians: An MPS point of view}
\title{Inner workings of fractional quantum Hall parent Hamiltonians:
A matrix product state point of view}

\author{Matheus Schossler}
\email{mschossler@wustl.edu}
\affiliation{Department of Physics, Washington University in St. Louis, 1 Brookings Dr., St. Louis MO 63130, USA}

\author{Sumanta Bandyopadhyay}
\affiliation{Nordita, KTH Royal Institute of Technology and Stockholm University, Roslagstullsbacken 23, SE-106 91 Stockholm, Sweden}

\author{Alexander Seidel}
\email{seidel@physics.wustl.edu}
\affiliation{Department of Physics, Washington University in St. Louis, 1 Brookings Dr., St. Louis MO 63130, USA}

\date{\today}

\begin{abstract}
We study frustration-free Hamiltonians of fractional quantum Hall (FQH) states from the point of view of the matrix product state (MPS) representation of their ground and excited states. There is a wealth of solvable models relating to FQH physics, which, however, is mostly derived and analyzed from the vantage point of first-quantized ''analytic clustering properties''. In contrast, one obtains long-ranged frustration-free lattice models when these Hamiltonians are studied in an orbital basis, which is the natural basis for the MPS representation of FQH states. The connection between MPS-like states and frustration-free parent Hamiltonians is the central guiding principle in the construction of solvable lattice models, but thus far, only for short-range Hamiltonians and MPSs of finite bond dimension. The situation in the FQH context is fundamentally different. Here we expose the direct link between the infinite-bond-dimension MPS structure of Laughlin-conformal field theory (CFT) states and their parent Hamiltonians. While focusing on the Laughlin state, generalizations to other CFT-MPSs will become transparent.
\end{abstract}
\maketitle
\section{Introduction }
The study of the fractional quantum Hall (FQH) effect is known for its bootstrap approach of producing phase diagrams via beautiful many-body wave functions \cite{Laughlin1983,Moore1991,Read1999,Jain1989,Halperin1983,haldane1988,Simon2007} that are pulled apparently out of thin air and linked to effective field theory using a sophisticated array of methods.
This seemingly bypasses the traditional Hamiltonian starting point of condensed matter physics. 
In truth, many of the most preeminent FQH trial wave functions admit \cite{Haldane1983,Trugman1985, Prange1990, Greiter1992, Halperin1983, Read1999, Simon2007, Greiter2021, Bandyopadhyay2020,simon_chapter_2021} parent Hamiltonians with very useful properties.
Such Hamiltonians do not only cement the status of certain states as incompressibile representatives of viable phases in the FQH regime. On top of that, they also identify a tower of zero energy (zero mode) states that describes both quasi-hole and edge degrees of freedom, and whose angular momentum spectrum is in one-to-one correspondence with the spectrum of a conformal field theory (CFT). This represents the almost ideal scenario of unambiguously extracting low energy physics from microscopic principles. Arguably, there is no other regime of correlated electron physics beyond one spatial dimension with similar analytic control.

At the same time, one may argue that the picture is not quite complete. 
Unlike in one dimension, where this is possible for many special Hamiltonians, there still lacks an analytic proof of the existence of an energy gap, a most defining feature of all of quantum Hall physics, but one that is not shared by all model Hamiltonians constructed in this context.\cite{haldane1988, Weerasinghe2014a}
%This is so even in cases where numerically, such a gap is most robust.
Indeed, as traditionally constructed, the FQH model Hamiltonians bear little resemblance with solvable models in one dimension. Typically, they are formulated as ultra-short ranged $k$-body interactions in the presence of some form of Landau level (LL) projection.
This description makes efficient use of 
analytic clustering properties of prototypical trial wave functions.
On the other hand, it leaves LL projection quite implicit, masking the physical degrees of freedom. \cite{Haldane2011} This is not ideal for making statements about spectral properties beyond the zero mode space. A radical departure from this is the 
study of FQH Hamiltonians in the LL occupation number basis, i.e., in second quantization. \cite{Chen2015, Mazaheri2015, Chen2015, Chen2017, Chen2018b, Bandyopadhyay2018, Bandyopadhyay2020}
In this approach, FQH parent Hamiltonians are indeed formally cast as one dimensional lattice models, albeit, crucially, with interactions that are not strictly finite ranged. 
At the pure wave function level, the last decade has seen similar radically new directions, driven by the insight that FQH model wave functions are matrix product states (MPS). \cite{Dubail2012a,Zaletel2012b,Tong2016,Estienne2013c,Wu2015,Crepel2018,Kjall2018}
The MPS description likewise tends to lead to a description in the occupation number basis.
Naively, these developments should elevate our understanding of FQH parent Hamiltonians to a level similar to that of traditional frustration free lattice models in one dimension. However, the infinite ranged nature of these lattice Hamiltonians, and the undoubtedly related fact that no finite bond dimension is associated to the corresponding FQH-MPS, represent considerably obstacles. Indeed, here we argue that the following may be the decisive difference between traditional frustration free 1D lattice models whose 
spectral feature are well understood, and the FQH model Hamiltonians in question: 
The pertinent 1D framework \cite{fannes1992} heavily capitalizes on the fact that the existence of
frustration free lattice models is understood {\em as a consequence} of the MPS stucture of the respective states  they stabilize. In other words, traditional frustration free 1D lattice models are constucted to capitalize on the MPS structure of certain wave functions, whereas quantum Hall model Hamiltonians are constructed to capitalize on analytic clustering properties of first quantized wave functions. So far, the only exception to the latter statement may be the composite fermion state Hamiltonians constructed in Ref. \onlinecite{Bandyopadhyay2020}, although this construction also did not make contact with an MPS. To the best of our knowledge, the existence of frustration free parent Hamiltonian in the FQH regime has thus far not been demonstrated using an orbital basis MPS description. 

In this paper, we intend to  develop such an angle of attack. 
Starting from a CFT, and the resulting MPS as constructed in the literature, we will demonstrate the frustration free character of the second quantized, ``lattice'' parent Hamiltonian directly from the orbital-basis MPS formulation of its zero modes.
While focusing on Laughlin states as the key example, the genericity of most our reasoning will be evident.

The remainder of this paper is organized as follows. In Sec. \ref{MPS} we 
review the MPS formalism for FQH states in the context of the Laughlin state.
In Sec. \ref{Ham} we review the second quantized representation of fractional quantum Hall parent Hamiltonians.
In Sec. \ref{induct} we present the induction step that is the heart of our formalism. It can be seen to be quite readily generalizable. 
Sec. \ref{indbegin} discusses those aspects that are, on the other hand, truly sensitive to details of the Hamiltonian and the CFT in questions. Formally, this is the induction beginning of our method.
In Sec. \ref{complete}, we discuss generic methods to establish the completness of a given class of CFT-MPS within the zero mode space of the corresponding parent Hamiltonian.
In Sec. \ref{QP} we take a detour and discuss variational (non-eigen) quasiparticle states that emerge naturally from the action of second quantized operator algebras on the MPS.
We conclude in Sec. \ref{conclusion}. \vspace{-0.5cm}
\section{MPS representation of Laughlin states \label{MPS}}
Being the most fundamental fractional quantum Hall state, the Laughlin state also has the simplest matrix product representation. 
Before we can elaborate on its relation to the 
existence of a local parent Hamiltonian, 
we first review the derivation of this MPS representation from CFT. \cite{Dubail2012a,Zaletel2012b,Estienne2013c,Crepel2018}
 The filling factor $\nu=1/q$ Laughlin state \cite{Laughlin1983}
for a system of $N$ particles has the wave function
\setlength{\belowdisplayskip}{5pt} \setlength{\belowdisplayshortskip}{5pt}
\setlength{\abovedisplayskip}{5pt} \setlength{\abovedisplayshortskip}{5pt}
\begin{equation}
\psi_{L}\left(z_{1}\cdots z_{N}\right)=\prod_{i<j}\left(z_{i}-z_{j}\right)^{q},\label{laughlin}
\end{equation}
where we suppress the exponential factor
$\exp[-\frac 14 \sum_i |z_i|^2]$ and focus on the polynomial part. 
%The exponential factor is dropped
%because it can be recovered through an appropriate choice of the background
%charge in the CFT. 
As is well-established in the literature \cite{FUBINI1991,Cristofano1991,Moore1991,di1996conformal,mussardo2009statistical},
and again well-covered recently with MPS in mind \cite{Zaletel2012b,Tong2016,Estienne2013c,Wu2015,Crepel2018,Kjall2018},
\Eq{laughlin}
can be written as the CFT correlator of a chiral free massless bosonic
theory, or ``Coulomb gas'', 
\begin{equation}
\psi_{L}\left(z_{1}\cdots z_{N}\right)=\left\langle V_{\sqrt{q}}\left(z_{N}\right)\cdots V_{\sqrt{q}}\left(z_{1}\right)\right\rangle ,\label{statate_correlator}
\end{equation}
where 
$V_{\sqrt{q}}\left(z\right)=:e^{i\sqrt{q}\,\phi\left(z\right)}:$
is the holomorphic (chiral) part of the vertex operator. More specifically, it is
a primary field in a chiral free massless bosonic  CFT in $1+1d$ with
$U(1)$ charge $\sqrt{q}$ (see below). 
Here we will only summarize defining properties of this theory and refer the interested reader to Ref. \onlinecite{di1996conformal,mussardo2009statistical}.
The chiral bosonic
field can be given the mode expansion defined by  \vspace{-0.2cm}
\begin{equation}\label{phimode}
\phi\left(z\right)=\phi_{0}-ia_{0}\log\left(z\right)+i\sum_{n\neq0}\frac{1}{n}a_{n}z^{-n},
\end{equation}
where the $a_{n}\text{'s}$ obey the algebra 
\begin{align}
\left[\phi_{0},a_{0}\right] =i,\quad 
\left[a_{n},a_{m}\right] =n\delta_{n+m,0}.
\end{align}
These operators represent neutral excitations of the %Coulomb gas
CFT. They act on states as follows: 
\begin{align}
a_{n}\ket{N} =0,\quad n>0, \quad
a_{0}\ket{N} =\sqrt{q}N\ket{N},\label{a_0psiN}
\end{align}
where $\ket{N}$ is a primary state of the bosonic theory. These states
together with the descendant states, 
\[\label{desc}
\left\{ a_{-n_{1}}a_{-n_{2}}\cdots\ket{N}\right\} ,\ \ \ n_{i}\in\mathbb{N},
\]
form a basis of the Fock space of this theory. The $U(1)$ charge $\sqrt{q}N$, measured by $a_0$ and shifted by $\exp(i\sqrt{q}\phi_0)$, will also play the role of being proportional to the particle number in the quantum Hall state to be constructed.
%
%For now, $N$ is just
%labeling an arbitrary state, but it %will become clear this is
%the number of particles in the %system. 
The $a_{n}\text{'s}$
operators are the raising operators that generate the neutral excitations of the chiral
free massless bosonic CFT.

The following steps now result in an MPS-form for the
Laughlin state Eq. (\ref{statate_correlator}). The vertex operator
admits the mode expansion,  \vspace{-0.2cm}
\begin{equation}
V_{\sqrt{q}}\left(z\right)=\sum_{\lambda}V_{-\lambda-h}z^{\lambda},\label{vertex_mode_exp}
\end{equation}
where $h=q/2$ is the conformal dimension of $V$ and 
\begin{equation}
V_{-\lambda}=\ointop\frac{dz}{2\pi i}z^{-\lambda-1}V_{\sqrt{q}}\left(z\right).\label{mode_def}
\end{equation}
The $V_{\sqrt{q}}\left(z\right)$ operator and its modes commute
(anti-commute) for $q$ even (odd), i.e., when the Jastrow factor in \Eq{laughlin} is describing a symmetric (anti-symmetric)
function. %
\begin{comment}
which corresponds to bosonic (fermionic) particle system in this case. 
\end{comment}
This mode expansion allows one to write the correlators \eqref{statate_correlator} of $V'$s
in terms of powers of $z'$s: 
\begin{align}
\psi_{N}\left(z_{1}\!\cdots\! z_{N}\right)\!=\!\! & \sum_{\left\{ \lambda_{i}\right\} }\!\!\braket{\alpha_{\text{out}}|V_{-\lambda_{N}-h}\!\cdots\! V_{-\lambda_{1}-h}|\alpha_{\text{in}}}\!\prod_{i=1}^{N}z_{i}^{\lambda_{i}}.\label{exp_correlator}
%\nonumber \\ & \times
\end{align}
Eq. (\ref{vertex_mode_exp}) formally includes all powers (positive
and negative) of $z$, but will only contain non-negative powers for the appropriate choice of ``out'' and ``in'' states $\bra{\alpha_{\text{out}}}$ and $\ket{\alpha_{\text{in}}}$, respectively, which we will now discuss. 
This will lead to the introduction of the background charge operator. %As we will see, the $\bra{\alpha_{\text{out}}}$ state sets the ``maximum'' value of the angular momentum, $\lambda_{\text{max}}=q(N-1)$, of the Laughlin state, i.e., the angular momentum after action with the full operator product, which will equal the angular momentum of the resulting MPS. 
As we will see, the $\bra{\alpha_{\text{out}}}$ state sets the ``maximum'' value of the angular momentum, $\lambda_{\text{max}}=q(N-1)$, of the Laughlin state, i.e., the largest angular momentum that will appear 
in the resulting orbital basis MPS. 
Correspondingly, we choose the $\ket{\alpha_{\text{in}}}$ state to have  ``minimum'' angular momentum $\lambda_{\text{min}}=0$. 
\begin{comment}
This result can be achieved using integral form of the $V_{-\lambda}$,
i.e., eq. (\ref{mode_def}), the normal ordering expansion (\ref{normal_ordering_expantion_V}),
and the properties (\ref{a_0psiN}) of $a_{n}$'s. 
\end{comment}
\begin{comment}
The second exponent generates all possible descendant states on top
of a primary state with no charge $(N=0)$. The exponential with $a_{0}$
is $1$ because $a_{0}$ reads zero from zero $U(1)$ charge (vacuum)
initial state $\ket{\alpha_{\text{in}}=0}$. The term with $\phi_{0}$ adds
one electron charge to this state. The resulting states has one electron
charge and all the infinity possible combinations of descendant states.
These states still have a $z$ dependence, in fact a combination of
positive powers of $z$. Finally, the contour integral of (\ref{mode_def})
returns identically zero for $\lambda<0$. In summary 
\end{comment}
%This property can be seen as a consistency requirement so the $\ket{\alpha_{\text{in}}}$
%and $\bra{\alpha_{\text{out}}}$ are well-defined. 
More concretely, the in-state $\ket{\alpha_{\text{in}}}$
is defined by the action on the vacuum $\ket{0}$ of the field $V_{\alpha_{\text{in}}\sqrt{q}}$ at the origin, 
\begin{align}
\ket{\alpha_{\text{in}}} & =\lim_{z\rightarrow0}V_{\alpha_{\text{in}}\sqrt{q}}\left(z\right)\ket{0}=e^{i\alpha_{\text{in}}\sqrt{q}\phi_{0}}\ket{0},
%\nonumber \\ & 
\end{align}
where the modes have the property $V_{-\lambda-h}\ket{\alpha_{\text{in}}}=0$  for $\lambda < q\alpha_{\text{in}}$, 
to ensure the absence of singularities \cite{di1996conformal,mussardo2009statistical}. Thus, $\alpha_{\text{in}}=0$ indeed leads to $\lambda_{\text{min}}=0$.
%where the modes have the property $V_{-\lambda-h}\ket{\alpha_{\text{in}}}=0$ for $\lambda<q\alpha_{\text{in}}$, which avoids singularities.
On the other hand,
 the out-state is defined
via the following limit,
\begin{align}\label{aR}
\bra{\alpha_{\text{out}}} & =\lim_{z\rightarrow\infty}z^{q\alpha_{\text{out}}}\bra{0}V_{-\alpha_{\text{out}}\sqrt{q}}\left(z\right)%\nonumber \\& 
=\bra{0}e^{-i\sqrt{q}\alpha_{\text{out}}\,\phi_{0}},
\end{align}
where $\bra{\alpha_{\text{out}}}V_{-\lambda-h}=0$ for $\lambda > q\left(\alpha_{\text{out}} -1 \right)$, i.e., $\lambda_\text{max}=q(\alpha_{\text{out}}-1)$. Given the above, and the number of vertex operators in (\ref{statate_correlator}),
we must now enforce
the neutrality condition for the bosonic CFT, which asserts that $-\alpha_{\text{out}}\sqrt{q}+N\sqrt{q}+\alpha_{\text{in}}\sqrt{q}=0$.
This condition must be satisfied, otherwise the correlator vanishes.
Since $\alpha_{\text{in}}=0$, we get $\alpha_{\text{out}}=N$. This is indeed consistent with $\lambda_\text{max}=q(N-1)$ being the highest occupied single particle angular momentum in the Laughlin state. Through \Eq{aR}, this introduces
 a background charge operator, 
\begin{equation}
O_{bg}=e^{-i\sqrt{q}N\,\phi_{0}},\label{Obg}
\end{equation}
to be added in the CFT correlator, which enforces the neutrality
condition. This operator is closely related to the one generating a uniform charge distribution, $\tilde O_{bg}=\exp\left(-i\rho\sqrt{q}\int d^{2}z\,\phi\left(z\right)\right)$,
where $\rho$ is the electric charge density. 
The only difference between these two operators only lies in the log-term in \Eq{phimode}.
This term generates the
Gaussian factor of the Laughlin wave function, modulo an everywhere singular gauge transformation \cite{Moore1991,Hansson2007,Zaletel2012b,Wu2015}. As is customary, we will usually drop this Gaussian factor, which is merely a spectator in all of the following. We will thus achieve charge neutrality through the operator $O_{bg}$ in Eqs. \eqref{aR}, \eqref{Obg}.

The Laughlin wave function (\ref{laughlin}) is symmetric (anti-symmetric)
for $q$ even (odd). This is also manifest in Eq. (\ref{exp_correlator}) by the (anti-)commutation of the modes $V_{-\lambda-h}$.
It is somewhat customary in the literature, though not strictly necessary for our purposes, to 
restrict sums over $\{\lambda_i\}$ by passing from 
un-ordered sequences $\{\lambda_i\}$ to ordered sequences $(\lambda_i)$ via
%(requires a factor $N!/\prod_{i}l_{i}!$), %
\begin{comment}
\begin{align}
0 & \leq\lambda_{1}\leq\cdots\lambda_{N}\leq qN\ \ \text{for bosons,}\\
0 & \leq\lambda_{1}<\cdots\lambda_{N}\leq qN\ \ \text{for fermions},
\end{align}
\end{comment}
\begin{align}
\left(\lambda_{i}\right): & \begin{cases}
0\leq\lambda_{1}\leq\cdots\lambda_{N}\leq \lambda_\text{max} & \text{for bosons,}\\
0\leq\lambda_{1}<\cdots\lambda_{N}\leq \lambda_\text{max} & \text{for fermions},
\end{cases} \label{def_ordered_set}
\end{align}
i.e., we now wish to restrict 
the sum in Eq. \eqref{exp_correlator} 
to only the ordered sets $(\lambda_i)$.
%which means dropping all terms where the exponent of any $z_{i}$ is different
%of $\lambda_{i}$.
Indeed, using the (anti-)symmetric commutators of the modes of the vertex operator, we may equivalently write\vspace{-0.2cm}
%The symmetric (anti-symmetric) 
%many particle function is then %recovered
%by symmetrizing 
%(anti-symmetrizing) resulting %expression in $\left\{ z\right\} $,
%{\em provided that}
%we introduce combinatorial %factors
%$N!/\prod_{\lambda}l_{\lambda}!$
%with each product of $z_i$'s,
\begin{align}
\psi_{N}\left(z_{1}\cdots z_{N}\right) & =\sum_{\left(\lambda_{i}\right)}\braket{\alpha_{\text{out}}|V_{-\lambda_{N}-h}\cdots V_{-\lambda_{1}-h}|\alpha_{\text{in}}}\nonumber \\
 & \times\frac{N!}{\prod_{\lambda}l_{\lambda}!}\frac{1}{N!}\sum_{\sigma\in S_{N}}\left({\rm sgn}\,\sigma\right)^{q}\prod_{i=1}^{N}z_{i}^{\lambda_{\sigma_i}},\label{preMPS}
\end{align}
where $l_\lambda$ is the number of occurrences of $\lambda$ among the $\lambda_i$. The ordering of the modes in equation (\ref{preMPS}) must be consistent with (\ref{def_ordered_set}) because of their (anti-)commutation relations. Note that $l_\lambda=0,1$ for fermions only. 

The second line in the last equations can be identified as an occupation number eigenstate
\begin{comment}
of the Fock space 
\end{comment}
 in the angular momentum basis of the Fock space. As a consequence, the last expression is manifestly the second quantized
description of the Laughlin state in this basis, 
\begin{equation}
\ket{\psi_{N}}=\sum_{\left(\lambda_{i}\right)}\braket{\alpha_{\text{out}}|V_{-\lambda_{N}-h}\cdots V_{-\lambda_{1}-h}|\alpha_{\text{in}}}\ket{\left(\lambda_{i}\right)},\label{ketN}
\end{equation}
where 
\begin{equation}
\braket{z_{1},\!\cdots\!,z_{N}|\left(\lambda_{i}\right)}\!=\!\frac{1}{N!\prod_{\lambda}l_{\lambda}!}\sum_{\sigma\in S_{N}}\left({\rm sgn}\,\sigma\right)^{q}\prod_{i=1}^{N}z_{i}^{\lambda_{\sigma_i}},\label{slater}
\end{equation}
\begin{comment}
\begin{equation}
\ket{\psi_{N}}=\sum_{\left(\lambda_{i}\right)}\frac{N!}{\prod_{i}l_{i}!}\braket{\alpha_{\text{out}}|V_{-\lambda_{N}-h}\cdots V_{-\lambda_{1}-h}|\alpha_{\text{in}}}\ket{\left(\lambda_{i}\right)},\label{ketN}
\end{equation}
where 
\begin{equation}
\braket{z_{1},\cdots,z_{N}|\left(\lambda_{i}\right)}=\frac{1}{N!}\sum_{\sigma\in S_{N}}\left({\rm sgn}\,\sigma\right)^{q}\prod_{i=1}^{N}z_{i}^{\lambda_{\sigma_{i}}},
\end{equation}
\end{comment}
%[HERE] The e
and, in \eqref{ketN}, \eqref{slater}, we dropped an overall normalization factor $N!$ present in \eqref{preMPS}, for convenience.
We point out that as a basis of the Fock space, the last equation is not only lacking the Gaussian factor, which is a trivial $(\lambda_i)$-independent multiplicative factor, but is in any case not normalized (even with the Gaussian factor in place). This is to say, we may think of this basis as created by pseudo-fermion  operators $ c_\lambda$, $ c_\lambda^\ast$, or their bosonic counterparts, where $ c_\lambda^\ast$ creates a particle in the un-normalized state
$z^\lambda \exp(-|z|^2/4)$,
and the associated destruction operators $c_\lambda$ may be defined 
such that $[ c_\lambda, c_{\lambda'}^\ast]_{\pm}=\delta_{\lambda,\lambda'}$ holds,
as in Ref. \onlinecite{Bandyopadhyay2020} (with upper [lower] sign for fermions [bosons]). We use this basis for convenience, however,
it is trivial to convert the coefficients in the above expansion to ones referring to a normalized basis, where the $\lambda$-dependence of proper single-particle normalization factors is given by $({2^{\lambda}\lambda!})^{-1/2}$. 
In the following, we will
for simplicity refer to 
the operators $c_\lambda, c_{\lambda'}^\ast$ as pseudo-particle operators, referring to both the fermionic and bosonic case on equal footing. They differ from ordinary  ladder operators chiefly through the fact that the Hermitian adjoint $(c_\lambda)^\dagger$  is only {\em proportional} to 
$c_\lambda^\ast$. The particle number operator can, however, be expressed as
$\hat N=\sum_\lambda c^\ast _\lambda c_\lambda$. We refer the reader to Ref. \onlinecite{Bandyopadhyay2020} for details.

With these remarks, \Eq{ketN} defines an MPS representation for the $N-particle$ Laughlin state in the basis $\ket{(\lambda_i)}$. \cite{Zaletel2012b,Tong2016,Estienne2013c,Wu2015,Crepel2018,Kjall2018}
 It is common in the literature \cite{Estienne2013c} to further represent the MPS coefficients, $\braket{\alpha_{\text{out}}|V_{-\lambda_{N}-h}\cdots V_{-\lambda_{1}-h}|\alpha_{\text{in}}}$, in terms of the ``site dependent'' matrices $A^{l_{\lambda}}\left[\lambda\right]\propto\left(V_{-\lambda-h}\right)^{l_{\lambda}}$: $\braket{\alpha_{\text{out}}|A^{l_{qN}}\left[qN\right]\cdots A^{l_{0}}\left[0\right]|\alpha_{\text{in}}}$, passing to a ``site-dependent MPS'' 
 with basis states labeled by orbital occupancies $l_\lambda$, thus trading the ``external labels'' $\lambda_i$ that list the occupied angular momenta for new labels $l_\lambda$, which list the occupancies of all possible angular momenta. While one may clearly switch back and forth between these representations, it will be advantageous for us to work with the original MPS carrying $\lambda_i$-labels, \Eq{ketN}.

In the present work, we want to make direct contact between the MPS representation of the Laughlin state \eqref{ketN} and the existence of a frustration free parent Hamiltonian. 
Such parent Hamiltonians in the fractional quantum Hall regime traditionally have the benefit of not only counting the incompressible ground state in question among their zero  modes, but also quasi-hole type excitations, which increase the angular momentum of the state. This increase is infinitesimal if the quasihole approaches the edge of the system, in which case the zero mode describes a gapless edge excitation. Though strictly speaking non-dispersing,  the physical character of these gapless edge modes becomes apparent when a confining potential is added, which may be taken to be proportional to the total angular momentum. In this way, one recovers gapless, linearly dispersing edge modes, whose mode counting one expects to be in one-to-one correspondence with the counting of modes in the CFT from which the incompressible ground state is constructed in the manner of \Eq{ketN}.
Indeed, one can construct variational 
MPS states for a set of edge excitations in a natural manner (see below). 
It will be an added benefit of this work that a closed formalism will emerge to establish these MPS-edge modes as a complete set of zero modes of some parent Hamiltonian. In this way, it will become most manifest that 
zero mode counting, which has a long tradition \cite{Read1996,Ardonne2001,Ardonne2002a,Seidel2011} in the field,
is in one-to-one correspondence with mode counting in the associated CFT. This ``zero mode paradigm'' is indeed what one expects of a ``good'' parent Hamiltonian.

We proceed by reviewing how variational edge modes naturally emerge in the MPS formalism. %All that is required is 
%the addition of neutral excitations of the coulomb gas CFT in the
%final state.\cite{some_relevant_paper}%
\begin{comment}
First let us discuss the state $\ket{\psi_{N}^{p_{n_{1}}p_{n_{2}}\cdots}}$.
The upper index's represents zero-energy excitations in the Laughlin
state $\ket{\psi_{N}^{p_{n_{1}}p_{n_{2}}\cdots}}=\frac{p_{n_{1}}}{q^{-1/2}}\frac{p_{n_{2}}}{q^{-1/2}}\cdots\ket{\psi_{N}}.$ 
\end{comment} 
In the conformal edge theory, a complete set of edge excitations at fixed particle number is generated by
products of the mode operators $a_{-n}$,
as in \Eq{desc}. It is thus natural, from a variational point of view, to replace $\ket{\alpha_{\text{out}}}=\ket{N}$ with \Eq{desc} in the MPS
\eqref{exp_correlator} in order to describe edge excitations.
This leads to the following form:
\begin{equation}
\ket{\psi_{N}^{a_{n}...}}=\sum_{\left(\lambda_{i}\right)}\braket{N|a_{n}\cdots V_{-\lambda_{N}-h}\cdots V_{-\lambda_{1}-h}|0}\ket{\left(\lambda_{i}\right)},\label{ketN_final}
\end{equation}
where $a_n\dotsc$
is short for any products of $M$ mode operators
$a_{n_1}a_{n_2}\dotsc a_{n_M}$. 
%The index set $\left(\lambda_{i}\right)$ is
%defined in equation (\ref{def_ordered_set}).
While the above motivates the expression \eqref{ketN_final}, it does not establish this expression from a Hamiltonian principle, which we will achieve below. As a stepping stone, it is worthwhile to motivate \Eq{ketN_final} in a slightly different manner.  

In first quantization,
it is well-known that a complete set of edge excitations of the Laughlin state can be generated 
by multiplication of the incompressible Laughlin state with 
general products of power-sum polynomials.
\cite{Wen1990,Stone1990,Read1996,Milovanovic1996,Mazaheri2015}
The equivalent 
of this operation in second quantization, which is more useful for present purposes, is given by the action of the operators
\begin{comment}
a short notation for the 
\end{comment}
 $p_{n}=\sum_{i=1}^{N}c_{\lambda_{i}+n}^{*}c_{\lambda_{i}}$,
 which facilitate multiplication
 with the power-sum polynomial $\sum_{k=1}^N z_k^n$. \cite{Mazaheri2015, Milovanovic1996}
 The effect of these operators on the Laughlin state in the MPS form \eqref{exp_correlator}
can be readily worked out:
\begin{align}
 & \sqrt{q}p_{n}\ket{\psi_{N}}\nonumber \\
 &\!\! =\!\!\sqrt{q}\!\sum_{\left(\lambda_{i}\right)}\!\!\braket{N|V_{-\lambda_{1}-h}\cdots V_{-\lambda_{N}-h}|0}\!\!\!\sum_{i=1}^{N}\!\ket{\lambda_{1},\cdots,\lambda_{i}+n,\cdots,\lambda_{N}}\nonumber \\
 &\! =\!\!\sqrt{q}\sum_{\left(\lambda_{i}\right)}\Bigl(\braket{N|V_{-\lambda_{1}+n-h}\cdots V_{-\lambda_{N}-h}|0}+\cdots\Bigr)\ket{\left\{ \lambda_{i}\right\} }\nonumber \\
 &\! =\!\!\sum_{\left(\lambda_{i}\right)}\braket{N|a_{n}V_{-\lambda_{1}-h}\cdots V_{-\lambda_{N}-h}|0}\ket{\left\{ \lambda_{i}\right\} }=\;\ket{\psi_{N}^{a_{n}}},\label{def_state_wth_excit}
\end{align}
where we used the commutation relation 
\begin{equation}
\left[a_{n},V_{-\lambda-h}\right]=\sqrt{q}\,V_{-\lambda+n-h},\label{comut_a_n,V_lambda}
\end{equation}
which follows from the more elementary CFT identity \cite{di1996conformal}
\begin{equation}
\left[a_{n},V_{\sqrt{q}}\left(z\right)\right]=\sqrt{q}z^{n}V_{\sqrt{q}}\left(z\right),
\end{equation}
together with Eq. (\ref{mode_def}) for the  modes of the vertex operator. $p_n$ adds $n$ to the Laughlin state total angular momentum and increases $\lambda_{\text{max}}$ from $q(\alpha_{\text{out}}-1)$ to $q(\alpha_{\text{out}}-1)+n$.
\begin{comment}
\begin{align}
\left[a_{n},V_{-\lambda-h}\right] & =\frac{1}{2\pi i}\ointop dzz^{-\lambda-1}\left[a_{n},V_{\sqrt{q}}\left(z\right)\right]\\
 & =\frac{\sqrt{q}}{2\pi i}\ointop dzz^{-\lambda+n-1}V_{\sqrt{q}}\left(z\right)\\
 & =\sqrt{q}V_{-\lambda+n-h}.
\end{align}
\end{comment}
\begin{comment}
\begin{align}
\left[a_{n},V_{-\lambda-h}\right] & =\frac{1}{2\pi i}\ointop dzz^{-\lambda-1}\left[a_{n},V_{\sqrt{q}}\left(z\right)\right]\\
 & =\sqrt{q}V_{-\lambda+n-h}.
\end{align}
\end{comment}
\Eq{def_state_wth_excit} can be iterated arbitrarily many times to give the
general edge excitation \eqref{ketN_final}.
% This now establishes these states as a complete set of zero modes {\em if} we already knew that the $p_n$ generate such when acting on $\ket{\psi_N}$. In the following section, we will re-derive this more rigorously.
This now establishes these states (in \ref{ketN_final}) as a complete set of zero modes in the MPS form, making a parallel with the power-sum polynomials in first quantization. In the following section, we will make the completeness of the set of zero modes in the MPS more transparent.
% \cite{Wen1990,Stone1990,Read1996,Milovanovic1996,Mazaheri2015}
This now establishes these states as a complete set of zero modes {\em if} we take as given that the $p_n$ generate such when acting on $\ket{\psi_N}$. While this can be shown in various ways, we will do so here without leaving the MPS framework.
%In the following section, we will re-derive this more rigorously.

%While other
%means exist to rigorously establish that
%the 
%states generated by the $p_n$ acting 
%on the Laughlin state represent a complete set of edge excitations relative to some parent Hamiltonian, both in first \cite{Stone1990} and second \cite{Mazaheri2015} quantization, 
%at present, the observation made in \Eq{def_state_wth_excit} does not change the variational character of these excited state. 
Moreover, the above establishes a one-to-one correspondence between operator algebras acting on the ``real'' Hilbert space describing microscopic degrees of freedom and modes representing the ``virtual'' degrees of freedom of the MPS. It is 
the fact that this kind of translation is
feasible that will make the developments of the following section possible, where we explore the action
of a Hamiltonian defined in terms of microscopic electron operators on the
MPS states \eqref{ketN_final}.
\section{The zero mode property}
\subsection{Hamiltonian and setup\label{Ham}}
We now derive the fact that Laughlin states admit parent Hamiltonians with the 
claimed features 
 from the MPS structure of the states \eqref{ketN_final}. To this end, we begin by reviewing pseudo-potentials in second quantization. 
From first quantization, it is well known that the $\nu=1/q$-Laughlin state is annihilated by 2-particle projection operators onto relative angular momentum $m$ known as Haldane pseudo-potentials \cite{Haldane1983}, with $m<q$
and $m$ even (odd) for $q$ even (odd). The second quantized version of such a Hamiltonian is of the general form \cite{Seidel2005,Ortiz2013a,Chen2015}
\begin{equation}
H_{\frac{1}{q}}=\sum_{\substack{0\leqslant m<q\\
\left(-1\right)^{m}=\left(-1\right)^{q}
}
}\sum_{R}Q_{R}^{m\dagger}Q_{R}^{m},\label{hamiltonian}
\end{equation}
where $Q^m_R$ is an operator that annihilates
a pair of particles with total angular momentum $2R$. The Hamiltonian is frustration free, in that the ground state is a zero mode of each 
of the (non-commuting) positive semi-definite terms $Q_{R}^{m\dagger}Q_{R}^{m}$.
The $Q_R^m$ can be chosen such that ${Q^m_R}^\dagger  Q^m_R$ is precisely the $m$th Haldane pseudo-potential. However, as we will be  interested only in the zero-mode space, and the zero mode condition for a ket $|\psi\rangle$ can be stated as
\begin{equation}
Q_{R}^{m}\ket{\psi}=0
\end{equation}
for all pertinent $m$ and $R$,
one may form linearly independent new linear combinations of the $Q_{R}^{m}$. The resulting Hamiltonian \eqref{hamiltonian} then has the same zero mode space as the sum of the original Haldane pseudo-potentials. With this in mind, it was established in \onlinecite{Ortiz2013a} that we may choose 
\begin{equation}
Q_{R}^{m}=\sum_{x}x^{m}c_{R-x} c_{R+x}.\label{def_Q}
\end{equation}
The above operators appear long-ranged in the distance $x$ only in pseudo-fermion/boson language. 
When the normalization factors are restored,  turning pseudo particles into ordinary electron destruction operators via $ c_k \rightarrow {\cal N}_k c_k$, the action of
\Eq{def_Q} is seen to be exponentially decaying in $x$. Indeed, for $m=0$ (bosons) and $m=1$ (fermions),
$\sum_R {Q^m_R}^\dagger Q^m_R $ with $Q^m_R$ chosen as in \Eq{def_Q} {\em is} (proportional to) the Haldane pseudo-potential with index $m$. While the relation is more complicated for larger $m$, all statements about zero modes that follow do equally apply to \Eq{def_Q} 
as well as the original Haldane pseudo-potentials parent Hamiltonian. Moreover, while we have the disk geometry in mind, all of the following applies also to the other genus zero geometries, i.e., the cylinder and sphere, where the Laughlin state is the same in terms of suitably defined pseudo-fermions (which is to say that its first quantized wave function is given by the same polynomial).

\begin{comment}
To study the zero mode property of the Laughlin wave function with
MPS approach we will start by learning how the Hamiltonian acts in
the angular momenta basis in the MPS form. The frustration-free parent
Hamiltonian for $\nu=1/q$ Laughlin state is a sum over the projections
of two particles with relative angular momentum less or equal than
$q$. \cite{Haldane1983,Simon2007a}%

Simon2007a,Haldane1983 

In second quantized language this Hamiltonian reads:%

\cite{Ortiz2013a,Chen2015} 

\begin{equation}
H_{\frac{1}{q}}=\sum_{\substack{0\leqslant m<q\\
\left(-1\right)^{m}=\left(-1\right)^{q}
}
}\sum_{R}Q_{R}^{m\dagger}Q_{R}^{m},\label{hamiltonian}
\end{equation}
where 
\begin{equation}
Q_{R}^{m}=\sum_{x}x^{m}c_{R-x}c_{R+x},\label{def_Q}
\end{equation}
with infinity radius cylinder form factor $x^{m}$.

The terms $Q_{R}^{m\dagger}Q_{R}^{m}$ acts as a projection onto a
subspace with relative angular momentum $2R\leq q$ of two particles
times a form factor 
\end{comment}
\begin{comment}
The later operator defines the zero mode condition: 
\[
Q_{R}^{m}\ket{\psi}=0,
\]
of the ground state of $H_{1/q}$ since it is a positive definite
operator. Because $Q_{R}^{m}$ can be extended to other genus zero
geometries (disk, sphere, finite radios cylinder) through a similarity
transformation \cite{Ortiz2013a}, which can be seen as taking a
linear combination of terms similar to (\ref{def_Q}), so it is sufficient
to work with (\ref{def_Q}).
\end{comment}

We are now in a position to demonstrate that the general MPS \eqref{ketN_final} is a zero mode of the Hamiltonian \eqref{hamiltonian}.
We do so by induction in particle number $N$. We present the induction step first. That is, we will first demonstrate
\begin{equation}
Q_{R}^{m}\ket{\psi_{N}^{a_{n}\cdots}}=0,\label{zero_mode_pro}
\end{equation}
for all half-integer $R\geq 0$, all $m$ included in the sum \eqref{hamiltonian}, and all strings $a_n\dotsc$, as long as the same statement is known for $N-1$ in place of $N$.
We note that this is a stronger induction assumption than that in Ref. \onlinecite{Chen2015}, which was used there to relate the zero mode property to a second quantized recursion relation related to Read's order parameter formalism \cite{Read1989} for the Laughlin state. Unlike in  Ref. \onlinecite{Chen2015}, we make an induction assumption for all zero mode, not just the incompressible Laughlin state. In this way, our reasoning becomes completely independent of 
particular properties of
the $p_n$ discussed earlier. This is the main reason why the present analysis can be generalized to other CFT-related quantum Hall states and their parent Hamiltonians.

The two body operator (\ref{def_Q}) obeys the following identity,
independent of the form factors (here, $x^m$) \cite{Chen2015}: 
\begin{equation}
Q_{R}^{m}\frac{1}{N}\sum_{\lambda\geq0}c_{\lambda}^{\ast}c_{\lambda}=\frac{2}{N}Q_{R}^{m}+\frac{1}{N}\sum_{\lambda\geq0}c_{\lambda}^{\ast}Q_{R}^{m}c_{\lambda}.\label{Q_ident}
\end{equation}
\begin{comment}
\begin{align}
Q_{R}^{m}\ket{\psi_{N}^{p_{n_{1}},p_{n_{2}},\cdots}} & =\frac{1}{N}\sum_{r\geq0}Q_{R}^{m}c_{r}^{\dagger}c_{r}\ket{\psi_{N}^{p_{n_{1}},p_{n_{2}},\cdots}}\\
 & =\frac{1}{N}\sum_{r\geq0}\sum_{x}\eta_{x}^{m}c_{R-x}c_{R+x}c_{r}^{\dagger}c_{r}\ket{\psi_{N}^{p_{n_{1}},p_{n_{2}},\cdots}}\\
 & =\frac{1}{N}\sum_{r\geq0}\sum_{x}\eta_{x}^{m}\left(\delta_{R+x,r}c_{R-x}c_{r}+\delta_{R-x,r}c_{r}c_{R+x}+c_{r}^{\dagger}c_{R-x}c_{R+x}c_{r}\right)\ket{\psi_{N}^{p_{n_{1}},p_{n_{2}},\cdots}}\\
 & =\left(\frac{2}{N}Q_{R}^{m}+\frac{1}{N}\sum_{r\geq0}c_{r}^{\dagger}Q_{R}^{m}c_{r}\right)\ket{\psi_{N}^{p_{n_{1}},p_{n_{2}},\cdots}},\nonumber 
\end{align}
\end{comment}
\begin{comment}
Here this relation is shown to be a direct consequence of the MPS structure
of the Laughlin state.%

With the MPS representation of a Laughlin state (\ref{ketN_final}), that includes the densest state and its zero-energy excitations, the zero
mode condition of the Hamiltonian (\ref{hamiltonian}) can be handled
through an induction proof in the spirit of \cite{Chen2015,Chen2018b}.

The induction consists in assuming that all states in this
class with $N-1$ particles with an arbitrary number of zero-energy
excitations on top of Laughlin state are zero modes, 
\begin{equation}
Q_{R}^{m}\ket{\psi_{N-1}^{p_{n_{1}},\cdots}}=0,\label{assumption_for_QpsiN-1}
\end{equation}
then proving the zero mode for all states with $N$ particles.
\end{comment}
The key insight in utilizing this equation is the realization that 
$c_\lambda$ acting on any $N$-particle state of the form
\eqref{ketN_final} yields a state to which the induction hypothesis applies, i.e., a linear combination of $N-1$ particle states of the form \eqref{ketN_final}. Then, when acting with
\Eq{Q_ident} on any state $\ket{\psi_{N}^{a_{n}\cdots}}$, the last term on the right vanishes by the induction assumption, and the remaining terms yield $(1-2/N)Q_{R}^{m}\ket{\psi_{N}^{a_{n}\cdots}}=0$,
therefore, \Eq{zero_mode_pro} follows if $N>2$.
%\frac{\sqrt{N}}{l_{\lambda}}
\subsection{The induction step\label{induct}}
To complete the induction step, we must therefore evaluate
\begin{align}
\sqrt{N}\,& c_{\!\lambda}\ket{\psi_{N}^{a_{n}\dotsc}} =\nonumber\\
\!\!\!\!\!\sum_{\left(\lambda_{i}\right)_{i=1\dotsc N-1}}&\!\!\!\!\!\!\!\!\!\braket{N|a_{n}\!\dotsc\! V_{\!-\lambda_{}-h}\!V_{\!-\lambda_{N-1}-h}\!\cdots\!V_{\!-\lambda_{1}-h}\!|0}\!\ket{\left(\lambda_{i}\right)_{i=1\dotsc N-1}}\!.\label{intial_cr_psiN}
\end{align}
In the above, it was used that
$\ket{\left(\lambda_{i}\right)_{i=1\dotsc N}}$ as defined in \Eq{slater} equals
$(\sqrt{N!} \prod_\lambda l_\lambda! )^{-1}c^\ast_{\lambda_1}\dotsc c^\ast_{\lambda_N}|0\rangle$, and that
the modes of $V_{\sqrt{q}}\left(z\right)$ commute (anti-commute)
for even (odd) $q$ so we can pull the mode corresponding to the annihilated angular
momentum $\lambda$ to the left.
For fermions, the sign generated by the anti-commutation of the modes of the vertex cancel exactly 
the sign due to the anti-commutation of the pseudo-fermion operators.
Let us, for simplicity, first consider the case where $a_n\dotsc=a_n$ is a single mode operator.
The relation (\ref{comut_a_n,V_lambda}) is used to apply
$V_{-\lambda-h}$
directly on $\bra{N}$, $\braket{N|a_{n}V_{-\lambda-h}V_{-\lambda_{N}-h}\cdots V_{-\lambda_{1}-h}|0}\rightarrow\braket{N|\left(V_{-\lambda-h}a_{n}+\sqrt{q}V_{-\lambda+n-h}\right)V_{-\lambda_{N}-h}\cdots V_{-\lambda_{1}-h}|0}$,
generating an extra term involving 
$V_{-\lambda-h+n}$ and
no $a_n$-mode.
The modes of the vertex operator
modify the final state in the following way: 
\begin{equation}
%\label{initial_bra_NtoN-1}
\bra{N}\!V_{-\lambda-h}\!\! =\!\bra{0}\!e^{-i\sqrt{q}N\!\phi_{0}}V_{\!-\lambda-h}\!=\bra{N-1}\!\!\!\!\!\!\!\!\sum_{l=0}^{q\left(N-1\right)-\lambda}\!\!\!\!\!\!\!\!b_{q\left(N-1\right)-\lambda}^{l}\label{bra_NtoN-1},
\end{equation}
where 
\begin{align}
b_{k}^{l} & =\frac{\left(-\sqrt{q}\right)^{l}}{l!}\!\!\!\!\!\!\sum_{i_{1}+\cdots+i_{l}=k}\!\!\frac{a_{i_{1}}}{i_{1}}\frac{a_{i_{2}}}{i_{2}}\cdots\frac{a_{i_{l}}}{i_{l}},\, i_{j}>0\Rightarrow k\geq l;\nonumber \\
b_{k}^{0} & =\begin{cases}
1,\! & k=0\\
0,\! & k>0.
\end{cases}\label{b_k^l_def-1}
\end{align}
The proof of this result will be left to Appendix A. The operators
$b_{k}^{l}$ are a collection of $l$ neutral excitations corresponding to the addition of angular momentum $k$ to the Laughlin state, i.e., they are in the algebra generated by the modes $a_n$, $n>0$. Inserting equation
(\ref{bra_NtoN-1}) into (\ref{intial_cr_psiN}) we find 
\begin{align}
&\sqrt{N} c_{\lambda}\ket{\psi_{N}^{a_{n}}}=\!\!\!\!\!\!\sum_{\left(\lambda_{i}\right)_{i=1\dotsc N-1}}\!\!\!\!\!\Bigl\langle\! N-1|\biggl(\sum_{l=0}^{q\left(N-1\right)-\lambda}\!\!\!\!\!b_{q\left(N-1\right)-\lambda}^{l}a_{n}+\nonumber \\
 & \sqrt{q}\!\!\!\!\!\!\!\!\sum_{l=0}^{q\left(N-1\right)-\lambda+n}\!\!\!\!\!\!\!b_{q\left(N-1\right)-\lambda+n}^{l}\biggr)V_{\!-\lambda_{N}-h}\!\cdots\! V_{\!-\lambda_{1}-h}|0\Bigr\rangle\!\ket{\left(\lambda_{i}\right)_{i=1\dotsc N-1}\!},
\end{align}
\begin{comment}
\begin{widetext}
\[
c_{\lambda_{r}}\ket{\psi_{N}^{a_{n}}}=\!\!\!\!\!\sum_{\lambda_{1},\cdots,\hat{\lambda_{r}},\cdots}\!\!\!\!\!\braket{N-1|\left(\sum_{l=0}^{q\left(N-1\right)-\lambda_{r}}\!\!\!\!\!b_{q\left(N-1\right)-\lambda_{r}}^{l}a_{n}+\sqrt{q}\sum_{l=0}^{q\left(N-1\right)-\lambda_{r}+n}\!\!\!\!\!b_{q\left(N-1\right)-\lambda_{r}+n}^{l}\right)V_{\!-\lambda_{N}-h}\!\cdots V_{\!-\lambda_{1}-h}|0}\ket{\lambda_{1},\!\cdots\!,\hat{\lambda_{r}},\!\cdots},
\]
\end{widetext}
\end{comment}
\begin{comment}
\begin{align}
 & c_{\lambda_{r}}\ket{\psi_{N}^{p_{n}}}=\sum_{\lambda_{1},\cdots,\hat{\lambda_{r}},\cdots}\nonumber \\
 & \times\bigl(\braket{N-1|\sum_{l=0}^{q\left(N-1\right)-\lambda_{r}}b_{q\left(N-1\right)-\lambda_{r}}^{l}a_{n}\cdots V_{-\lambda_{1}-h}|0}\ket{\lambda_{1},\cdots,\hat{\lambda_{r}},\cdots}+\nonumber \\
 & \sqrt{q}\braket{N-1|\sum_{l=0}^{q\left(N-1\right)-\lambda_{r}+n}b_{q\left(N-1\right)-\lambda_{r}+n}^{l}\cdots V_{-\lambda_{1}-h}|0}\ket{\lambda_{1},\cdots,\hat{\lambda_{r}},\cdots}\bigr),\label{anih_on densest}
\end{align}
\end{comment}
which, by the induction assumption, is a linear combination of $N-1$ particle zero modes. Therefore, the action
of $Q_{R}^{m}$ on $c_{\lambda}\ket{\psi_{N}^{a_{n}}}$ is zero.
%by the induction assumption (\ref{assumption_for_QpsiN-1}).
\begin{comment}
the Now that we know how to act an annihilation operator in the densest
state we can proceed with the demonstration of (\ref{zero_mode_pro}).we
will prove 
\[
Q_{R}^{m}c_{r}\ket{\psi_{N}^{p_{n_{1}},\cdots}}=0.
\]
\end{comment}
We can, of course, similarly treat $c_\lambda \ket{\psi_N}$.
Moreover, using the same method (and induction in $j$), we generally have that 
\begin{equation}
 \bra{N}a_{n_1} \dotsc a_{n_j} V_{-\lambda-h}
 =\bra{N-1}A\;,
\end{equation}
where $A$ is an element of the algebra generated by the $a_n$, $n>0$. The induction hypothesis then implies just as before that  $c_{\lambda}\ket{\psi_{N}^{a_{n_1}\dotsc a_{n_j} }}$ is a zero mode. This completes the induction step.
One may recognize the generality of these arguments, using only general properties of the operators $Q^m_R$, and, despite our limitation to the free chiral bosonic case, arguably the CFT in question. Indeed, the 
case of $k$-body generalization of the $Q^m_R$ is also quite analogous, the only difference being that the induction needs to start at $N=k$. 
Therefore, the only aspect that truly depends on details of the $Q^m_R$-like operators and on the CFT is the induction beginning. Below we continue discussing this for the Laughlin states.

\subsection{The zero mode property for two particles\label{indbegin}}

Now let us prove the zero mode property (\ref{zero_mode_pro}) for
$N=2$. In this case we can prove that a state with arbitrary number
of edge excitations can always be written as a linear combination
of
\begin{equation}\label{2p}
\ket{\psi_{2}^{n,l}}=\frac{1}{2}\sum_{\left\{ \lambda_{1},\lambda_{2}\right\} }\braket{2|V_{-\lambda_{2}+n-h}V_{-\lambda_{1}+l-h}|0}\ket{\lambda_{1},\lambda_{2}},
\end{equation}
\begin{comment}
\begin{equation}
\sum_{\left(\lambda_{i}\right)}\braket{2,0|V_{-\lambda_{2}+n}V_{-\lambda_{1}+l}|0}\ket{\lambda_{1},\lambda_{2}},
\end{equation}
\end{comment}
for arbitrary $n$ and $l$, where we have returned to the non-ordered
set $\left\{ \lambda_{1},\lambda_{2}\right\} $ with the addition of the factor $1/2$.
For two excitations $a_{l},\ l>0$, and $a_{n},\ n>0$,  we have, according to (\ref{comut_a_n,V_lambda}),
\begin{align}
  \frac{1}{q}\big\langle  a_{l}a_{n}\! &V_{-\lambda_{2}-h} V_{-\lambda_{1}-h} \big\rangle\! = \nonumber \\  
=&\phantom{+}\left\langle V_{-\lambda_{2}-h}V_{-\lambda_{1}+n+l-h}\!\right\rangle +\left\langle V_{-\lambda_{2}+l-h}V_{-\lambda_{1}+n-h}\!\right\rangle \nonumber \\
\ &+\left\langle V_{-\lambda_{2}+n-h}V_{-\lambda_{1}+l-h}\!\right\rangle +\left\langle V_{-\lambda_{2}+n+l-h}V_{-\lambda_{1}-h}\!\right\rangle .\label{two_particle_excit}
\end{align}
Adding more than two excitations to the state will only result in terms of the same general type, and, moreover, the case of zero and one excitations yield, of course, special instances thereof. 
Hence, it is enough to prove $Q^m_R\ket{\psi^{n,l}_2}=0$, % 
\begin{comment}
\begin{equation}
Q_{R}^{m}\ket{\psi_{2}^{p_{n},p_{l}}}=0,
\end{equation}
\end{comment}
with $Q_{R}^{m}$ given by (\ref{def_Q}), for non-negative integers
integer $n$ and $l$:%
\begin{align}
 &\braket{0|Q_{R}^{m}|\psi_{2}^{n,l}} =\!\!\!\!\sum_{x=-R}^{R}x^{m}\frac{1}{2}\!\!\sum_{\left\{ \lambda_{1},\lambda_{2}\right\} }\!\!\!\braket{2,0|V_{-\lambda_{2}+n-h}V_{-\lambda_{1}+l-h}|0}\nonumber \\
 & \times\left(\delta_{R+x,\lambda_{2}}\delta_{R-x,\lambda_{1}}+\left(-1\right)^{q}\delta_{R+x,\lambda_{1}}\delta_{R-x,\lambda_{2}}\right)\nonumber \\
 & =\delta_{q,2R-n-l}\!\!\!\sum_{x=-R}^{R}x^{m}\left(-1\right)^{R+x-n}{q \choose R+x-n}\nonumber \\
 &=\delta_{q,2R-n-l}\!\!\!\sum_{x'=-n}^{q+l}\left(x'-\frac{q-n+l}{2}\right)^{m}\!\!\!\left(-1\right)^{x'}{q \choose x'}=0
\end{align}
Here, the first line is found by acting with the operator $Q_{R}^{m}$
on the two-particle state \eqref{2p}. The second line is found using the result for the correlator of two modes of the vertex operator given in Appendix B. %
\begin{comment}
, then eliminating the sum over $\left(\lambda_{i}\right)$ by using
the two terms of product of Kronecker deltas
\end{comment}
\begin{comment}
by using the pair $\delta_{R+x,\lambda_{2}}\delta_{R-x,\lambda_{1}}$
\end{comment}
\begin{comment}
the correlator form of a Laughlin state of two particles to obtain
the third equation, the binomial expansion for the fourth equation,
and the definition of Kronecker delta for the last equation. From
the later one it implies $k=R+x-n,$ and $R=\frac{q+n+l}{2}$: 
\begin{align}
 & Q_{R}^{m}\ket{\psi_{2}^{p_{n},p_{l}}}=\\
 & =\delta_{q,2R-n-l}\sum_{x=-R}^{R}x^{m}\left(-1\right)^{R+x-n}{q \choose R+x-n}\ket{0}\\
 & =\delta_{q,2R-n-l}\sum_{x'=-n}^{q+l}\left(x'-\frac{q+l-n}{2}\right)^{m}\left(-1\right)^{x'}{q \choose x'}\ket{0}\\
 & =0,
\end{align}
\end{comment}
\begin{comment}
The third line we just used the restriction $R=(q+n+l)/2$ with a
change of variables
\end{comment}
The final line is found using the more elementary result $\sum_{0\leq j\leq q}\left(x-j\right)^{q}\left(-1\right)^{j}{q \choose j}=q!\ \forall x$, \cite{Ruiz1996} and differentiating it
with respect to $x$, finding that the sum vanishes  for all $m<q$, as expected. %
\begin{comment}
in the same way as in \cite{Chen2015} with the use of the first
Kronecker delta $R=(q+n+l)/2$
\end{comment}
This concludes our demonstration how the existence of frustration-free parent Hamiltonians for the Laughlin states can be seen as a consequence of their MPS-structure.
% \vspace{-0.3cm}
\subsection{Considerations of completeness\label{complete}}
Having established the existence of a parent Hamiltonian with zero modes \Eq{ketN_final}, we would also like to remark on how to establish the completeness of these zero modes without leaving the present framework.
It has been known for some time \cite{Rezayi1994, Bernevig2008} that 
the expansion of Laughlin states, and similarly other quantum Hall states, in Salter determinants $\ket{(\lambda_i)}$
or their bosonic equivalents is characterized by certain root states
$\ket{(\lambda_i^r)}$. The defining property 
of $\ket{(\lambda_i^r)}$ is that it has non-zero amplitude in the expansion of the underlying zero mode, and that all other $\ket{(\lambda_i)}$ with this property can be obtained from $\ket{(\lambda_i^r)}$ via sequences of  inward-squeezing processes $c_i^\ast c_j^\ast c_{j+r} c_{i-r}$, where $i<j$, $r>0$. The root state agrees with the thin torus limit \cite{Seidel2005,Seidel2006,Seidel2008a,Seidel2010,Bergholtz2007,Bergholtz2006b,Bergholtz2005,Rezayi1994}. Furthermore, $\ket{(\lambda_i^r)}$ is subject to rules known as generalized Pauli principles (GPPs) that are characteristic of the quantum Hall state and its quasi-hole excitations. In the case of the Laughlin state, the GPP dictates that there is no more than 1 particle in $q$ adjacent sites. The root state of the incompressible Laughlin state is simply the densest Slater determinant consistent with this rule, having occupancy $100100100100\dotsc$ for $q=3$.  

It is worth noting that these concepts have non-trivial generalizations to multicomponent and mixed-LL states \cite{Bandyopadhyay2018, Chen2017}, leading to entanglement at root level. The root state can then be thought of as the ``entangled DNA'' of the quantum Hall state, encoding much and more of its topological properties. In essence, the GPP becomes an ``entangled Pauli principle'' (EPP).

For both single Landau level and mixed-Landau level states, a general framework exists to derive the GPP/EPP as necessary conditions on root states of zero modes of the respective parent Hamiltonian of the general form \eqref{hamiltonian}, \cite{Ortiz2013a,Chen2015, Chen2017, Bandyopadhyay2018, Bandyopadhyay2020} assuming one exists.
(Again, generalization to $k$-particle operators is possible.) 
One consequence of this framework is that there cannot be more (linearly independent) zero modes than root states consistent with the GPP/EPP. Thus, whenever one has found one zero mode for every permissible root pattern, one is assured of the completeness of such a set of zero modes. This can be applied in the present situation: For MPS quantum Hall states, Estienne et. al \cite{Estienne2013c} have derived the systematics of extracting root states. This shows that the states \eqref{ketN_final} realize all possible root patterns consistent with the GPP, and are thus the complete set of zero modes for the Hamiltonian \eqref{hamiltonian}. 
% \vspace{-0.5cm}
\subsection{Quasi-electrons\label{QP}}

In principle, quasi-electron states are expected to emerge as finite energy eigenstates of the respective parent Hamiltonian.  Unfortunately, for all Hamiltonians known to us describing fractional quantum Hall states, such finite energy eingenstates have not been obtained exactly. Many variational constructs exist \cite{Laughlin1983, MacDonald1986, Jeon2003, Hansson2017, Bochniak2021a}. Here we elaborate on a construction that seems both natural to us and has a simple MPS representation. Consider again \Eq{def_state_wth_excit}. As we have seen,
the operators $p_n$ generate all possible zero modes by acting on the incompressible Laughlin state. Moreover, via the second quantized framework developed here, the action of such operators easily translates into MPS.
One natural approach is to consider the action of the adjoints, or, related to that (by form factors arising from the pseudo-particle nature of the $c_\lambda$, $c_\lambda^\ast$) the operators $p_{-n}$, $n>0$. A similar approach was taken recently in Ref. \onlinecite{Bochniak2021a}, where a quasi-electron operator has been constructed that exactly fractionalizes, in that $q$ applications exactly correspond to the local addition of one electron in the lowest Landau level, mirroring the celebrated property \cite{Girvin1984} of Laughlin's quasi-{\em hole} operator. Here we take a simplified approach, and consider the action of the $p_{-n}$ operator's on the incompressible Laughlin state. This lowers angular momentum, and so leads to states with nonzero energy in principle, at least for finite particle number. In complete analogy with \Eq{def_state_wth_excit}, one finds:
\begin{align}
 & \sqrt{q}p_{-n}\ket{\psi_{N}}  \nonumber  \\
 %& \frac{p_{-n}}{q^{-1/2}}\sum_{\left(\lambda_{i}\right)}\braket{N|V_{-\lambda_{1}-h}\cdots V_{-\lambda_{N}-h}|0}\ket{\lambda_{1},\cdots,\lambda_{i},\cdots,\lambda_{N}} \nonumber \\ 
= & \sqrt{q}\!\!\sum_{\left(\lambda_{i}\right)}\!\!\braket{N|V_{-\lambda_{1}-h}\!\cdots\! V_{-\lambda_{N}-h}|0}\!\!\!\sum_{i=1}^{N}\!\ket{\lambda_{1},\!\cdots\!,\lambda_{i}-n,\!\cdots\!,\lambda_{N}} \nonumber \\
%= & \sqrt{q}\sum_{\left(\lambda_{i}\right)}\Bigl(\braket{N|V_{-\lambda_{1}-n-h}\cdots V_{-\lambda_{N}-h}|0}+\cdots\Bigr)\ket{\left\{ \lambda_{i}\right\} }\label{to a_- on the right}  \nonumber \\
= & \sum_{\left(\lambda_{i}\right)}\braket{N|V_{-\lambda_{1}-h}\cdots V_{-\lambda_{N}-h}\,a_{-n}|0}\ket{\left\{ \lambda_{i}\right\} } \label{a_- on the right} =:  \ket{\psi_{N}^{a_{-n}}},
\end{align}
\begin{comment}
proof of equation (\ref{a_- on the right}) to (\ref{to a_- on the right}),
or in the other direction:
\begin{align}
 & V_{-\lambda_{N-1}-h}V_{-\lambda_{N}-h} a_{-n}=V_{-\lambda_{N-1}-h}\left(a_{-n}V_{-\lambda_{N}-h}+\sqrt{q}V_{-\lambda_{N}-n-h}\right)\\
 & =\cdots\left[V_{-\lambda_{N-1}-h}a_{-n}V_{-\lambda_{N}-h}+V_{-\lambda_{N-1}-h}\sqrt{q}V_{-\lambda_{N}-n-h}\right]\\
 & =\cdots\big[\left(a_{-n}V_{-\lambda_{N-1}-h}+\sqrt{q}V_{-\lambda_{N-1}-n-h}\right)V_{-\lambda_{N}-h} + \\
 &+V_{-\lambda_{N-1}-h}\sqrt{q}V_{-\lambda_{N}-n-h}\big]\\
 & =\cdots\big[a_{-n}V_{-\lambda_{N-1}-h}V_{-\lambda_{N}-h}+ \\ 
 & + \sqrt{q}V_{-\lambda_{N-1}-n-h}V_{-\lambda_{N}-h}+V_{-\lambda_{N-1}-h}\sqrt{q}V_{-\lambda_{N}-n-h}\big]
\end{align}
\end{comment}
with obvious generalization to the action of products of $p_{-n}$ operators. 
Having well-defined angular momentum, one can interpret the above as a basis for extended quasi-electron states, with the benefit of having {\em both} a nice MPS structure {\em and} a known generating principle in terms of second quantized electron operators. We expect that this basis may be key to understanding relations between different quasi-electrons constructions found in the literature, such as Ref. \onlinecite{Kjall2018,Hansson2017}, which emphasizes MPS formalism, and Ref. \onlinecite{Bochniak2021a}, which emphasizes second quantized operator algebras. 
%We note that the presence/absence of the geometric factors, introduced by the uniform background charge operator below eq. (\ref{Obg}), can make a substantial difference.
 We stress again that for quasi-holes and in the context of the Haldane pseudo-potential parent Hamiltonian, Eq. (\ref{def_state_wth_excit}) is an exact quasi-hole basis, whereas for quasi-particles, Eq. (\ref{a_- on the right}) is variational in character.
Detailed energetic considerations necessitate putting back the Gaussian factor dropped below \Eq{Obg}.\cite{Crepel2019e}

% \vspace{-0.5cm}

\section{Conclusion\label{conclusion}}

In this work we have explicitly linked this existence 
of a frustration free parent Hamiltonian
of the Laughlin states
to the MPS structure of these states.
In doing so, we discussed the action of certain second quantized operator algebras on CFT-derived MPS, which, in general, also lends itself to a discussion of quasi-electrons of {\em either} sign. 
While we leave details to future work \cite{Schosslera}, the framework is expected to generalize to many existing parent Hamiltonians in the FQH regime. We argued that this elevates the theory of these Hamiltonian to a level that is more on par with the theory of frustration free lattice models in one dimension. Specifically, FQH parent Hamiltonians have recently been constructed that are not obviously determined by clustering properties of first quantized wave functions \cite{Bandyopadhyay2020}, counter to the tradition of the field. It is, however, not obvious how to generalize the construction of Ref. \onlinecite{Bandyopadhyay2020}. The present construction is expected to make such generalization possible. We are hopeful that this will spur developments leading 
to a wealth of new solvable models.\vspace{-0.3cm}

\begin{acknowledgments}
\vspace{-0.3cm}
A.S. acknowledges insightful discussions with A. Bochniak, L. Chen, Z. Nussinov, G. Ortiz, and K. Yang.
This work has been supported
 by the National Science Foundation under Grant No. DMR-2029401.
\end{acknowledgments}

\begin{comment}

A natural sequence of this work is develop a similar scheme for non-abelian
states. The Moore-Read state is a strong candidate, since it is also
a free theory, but with fermionic degrees of freedom on its correlator.
In this theory the electron operator is a tensor product of the vertex
operator, same as in the bosonic correlator, and Majorana field. Although
it has different properties from the pure free bosonic theory, this
CFT theory is also explored in the literature \cite{di1996conformal,Nayak2008,Hansson2017}.

\end{comment}

% \bibliography{lib}
% \vspace{-0.1cm}
\section*{Appendix A}
% \vspace{-0.1cm}
\setlength{\belowdisplayskip}{4pt} \setlength{\belowdisplayshortskip}{4pt}
\setlength{\abovedisplayskip}{4pt} \setlength{\abovedisplayshortskip}{4pt}
In this appendix we will show the equation (\ref{bra_NtoN-1}). The
normal ordering expansion of the holomorphic part of the vertex operator is \vspace{-0.3cm}
\begin{equation}
V_{\sqrt{q}}\left(z\right)=e^{i\sqrt{q}\,\phi_{0}}e^{\sqrt{q}a_{0}\log\left(z\right)}\prod_{n=1}^{\infty}e^{\frac{\sqrt{q}}{n}\frac{a_{-n}}{z^{-n}}}e^{-\frac{\sqrt{q}}{n}\frac{a_{n}}{z^{n}}}.\label{normal_ordering_expantion_V}
\end{equation}
The action of the vertex operator in the dual state is then
\begin{align}
\bra{0} & e^{-i\sqrt{q}N\phi_{0}}V_{-\lambda-h}= \bra{0}e^{-i\sqrt{q}N\phi_{0}}\ointop\frac{dz}{2\pi i}\frac{V_{\sqrt{q}}\left(z\right)}{z^{\lambda+1}}\nonumber \\
= & \bra{N-1}\ointop\frac{dz}{2\pi i}z^{q\left(N-1\right)-\lambda-1}\sum_{l=0}^{\infty}\frac{1}{l!}\left(\sum_{n=1}^{\infty}\frac{-\sqrt{q}}{n}\frac{a_{n}}{z^{n}}\right)^{l},\label{intermediate_bra_NtoN-1}
\end{align}
where we have used the Taylor expansion 
\begin{equation}
e^{-\sqrt{q}\sum_{n=1}^{\infty}\frac{1}{n}\frac{a_{n}}{z^{n}}}=\sum_{l=0}^{\infty}\frac{1}{l!}\left(\sum_{n=1}^{\infty}\frac{-\sqrt{q}}{n}\frac{a_{n}}{z^{n}}\right)^{l}.
\end{equation}
Expanding again in powers of $z$ each term with power $l$ in the
above equation one finds \vspace{-0.2cm}
\begin{align}
\frac{1}{l!}\bigg(\sum_{n=1}^{\infty} &\frac{-\sqrt{q}}{n}a_{n}z^{-n}\bigg)^{l}=  \frac{1}{l!}\frac{\left(-\sqrt{q}a_{1}\right)^{l}}{z^{l}}+\frac{1}{l!}\frac{\left(-\sqrt{q}\right)^{l}}{z^{\left(l+1\right)}}\times\nonumber \\
 &\times\left(\sum_{i_{1}+\cdots+i_{l}=l+1}\frac{a_{i_{1}}}{i_{1}}\frac{a_{i_{2}}}{i_{2}}\cdots\frac{a_{i_{l}}}{i_{l}}\right) +\cdots = \sum_{k=l}^{\infty}b_{k}^{l}z^{-k},
\end{align}
\vspace{-0.1cm}with $b_{k}^{l}$ defined in (\ref{b_k^l_def-1}), yielding 
\begin{align}
\bra{N}V_{-\lambda-h} & =\bra{N-1}\sum_{\substack{l=0\\
k=l
}
}^{\infty}\sum_{k=l}^{\infty}b_{k}^{l}\ointop\frac{dz}{2\pi i}\,z^{q\left(N-1\right)-\lambda-1-k}\nonumber \\
% \end{align}\vspace{-0.8cm}
% \begin{align}
%  \phantom{\bra{N}V_{-\lambda-h}}
& =\bra{N-1}\sum_{l=0}^{\infty}\sum_{k=l}^{\infty}b_{k}^{l}\delta_{q\left(N-1\right)-\lambda-k,0}.\label{modes -> as}
\end{align}
The last sum can be evaluated using the definition of $b_{k}^{l}$
and the Kronecker delta. As a consequence equation (\ref{bra_NtoN-1})
is found.
\vspace{-0.7cm}
\section*{Appendix B}
\vspace{-0.3cm}
In this appendix we will calculate the correlator of two
modes of the vertex operator.\vspace{-0.2cm}%\vspace{-0.5cm}
\begin{comment}
\begin{align*}
 & =\sum_{x}x^{m}\frac{1}{\left(2\pi i\right)^{2}}\ointop\frac{dz_{1}}{z_{1}^{\left(R-x\right)-n+1}}\ointop\frac{dz_{2}}{z_{2}^{\left(R+x\right)-l+1}}\braket{2,0|V\left(z_{1}\right)V\left(z_{2}\right)|0}\ket{0}\\
 & =\sum_{x}x^{m}\frac{1}{\left(2\pi i\right)^{2}}\ointop\frac{dz_{1}}{z_{1}^{\left(R-x\right)-n+1}}\ointop\frac{dz_{2}}{z_{2}^{\left(R+x\right)-l+1}}\left(z_{1}-z_{2}\right)^{q}\ket{0}\\
 & =\sum_{x}x^{m}\sum_{k=0}^{q}\left(-1\right)^{k}{q \choose k}\ointop\frac{dz_{1}}{2\pi i}\frac{z_{1}^{q-k}}{z_{1}^{\left(R-x\right)-n+1}}\ointop\frac{dz_{2}}{2\pi i}\frac{z_{2}^{k}}{z_{2}^{\left(R+x\right)-l+1}}\ket{0}
\end{align*}
\end{comment}
\begin{align}
\braket{2,0|V_{-a-h}V_{-b-h}|0}\! & =\!\!\frac{1}{\left(2\pi i\right)^{2}}\!\!\ointop\!\!\!\frac{dz_{1}}{z_{1}^{a+1}}\!\!\ointop\!\!\!\frac{dz_{2}}{z_{2}^{b+1}}\!\!\braket{2|V\left(z_{1}\right)V\left(z_{2}\right)|0}\nonumber \\
 & =\!\frac{1}{\left(2\pi i\right)^{2}}\!\!\ointop\!\!\frac{dz_{1}}{z_{1}^{a+1}}\!\!\ointop\!\!\frac{dz_{2}}{z_{2}^{b+1}}\left(z_{1}-z_{2}\right)^{q}\nonumber \\
 & =\!\!\sum_{k=0}^{q}\!{q \choose k}\!\frac{\left(-1\right)^{k}}{\left(2\pi i\right)^{2}}\!\!\ointop\!\!\frac{dz_{1}}{2\pi i}\frac{z_{1}^{q-k}}{z_{1}^{a+1}}\!\!\ointop\!\!\frac{dz_{2}}{2\pi i}\frac{z_{2}^{k}}{z_{2}^{b+1}}\nonumber \\
 & =\!\!\sum_{k=0}^{q}\!\left(-1\right)^{k}\!{q \choose k}\delta_{q-k,a}\delta_{k,b}\! =\!\frac{\delta_{q,a+b}}{\left(-1\right)^{b}}\!{q \choose b}
\end{align}
where we have used the binomial expansion in the second line to find the third one.
\bibliographystyle{apsrev4-1}
\addcontentsline{toc}{section}{\refname}
% \bibliography{lib}

\end{document}